\documentclass[pre,twocolumn,secnumarabic]{revtex4-2}
\usepackage{graphicx} 
\usepackage{amsmath}
\usepackage{tikz-cd}
\usepackage{pgfplots}
\usetikzlibrary{arrows.meta}
\usetikzlibrary{external}
\usepackage{gensymb}
\newcommand{\up}{\uparrow}
\newcommand{\down}{\downarrow}

\begin{document}

\title{Using precision coefficients on recurrence times and integrated currents to lower bound the average dissipation rate}
\thanks{\copyright2026 American Physical Society. This is the accepted manuscript of the following article: Alberto Garilli and Diego Frezzato, “Using precision coefficients of recurrence times and integrated currents to construct a lower bound for the average dissipation rate,” Phys. Rev. E, Vol. 113, No. 5, 054127 (2026). The final published version is available from DOI:  https://doi.org/10.1103/gycp-9xg4.}
\author{Alberto Garilli}
\email{alberto.garilli@unipd.it}
\author{Diego Frezzato}
\affiliation{Department of Chemical Sciences, University of Padova, via Marzolo 1, I-35131, Padova, Italy.}

\date{\today}

\begin{abstract}
   For continuous-time Markov jump processes on irreducible networks with time-independent rate constants, we employ a transition-based formalism to express the long-time precision of a single integrated current over an observable channel in terms of precisions of the recurrence times of the forward and backward jumps, and of an effective affinity that captures the thermodynamic driving on that channel. This leads to a general inequality that, similarly to the well-known Thermodynamic Uncertainty Relation (TUR), links the stationary entropy production rate with the fluctuations of an integrated current, but also incorporates the statistics of the forward and backward recurrence times. Such inequality can be saturated in less restrictive conditions than the TUR, and potentially offers new opportunities for the optimization and design of biological and chemical out-of-equilibrium systems at the nanoscale.
\end{abstract}

\maketitle

\section{Introduction}
In small non-equilibrium systems, fluctuations play a crucial role and set fundamental constraints on performance. Stochastic thermodynamics provides the theoretical framework to quantify these limits, revealing how precision, speed, and energetic cost are tightly intertwined. A paradigmatic result in this direction is the Thermodynamic Uncertainty Relation (TUR) \cite{seifert_barato_TUR,gingrich2016dissipation,pietzonka2017finite,horowitz2017proof}, which provides a quantitative link between fluctuations and dissipation.

For Markov jump processes on irreducible networks and reversible jumps, in particular, the TUR concerns a general lower bound to the precision of any time-integrated current. In this work, we shall focus on the integrated steady current $\mathcal{N}_t = \mathcal{N}_{\alpha\to\beta}^t - \mathcal{N}_{\beta\to\alpha}^t$ along a channel connecting the pair of states $(\alpha,\beta)$, where $\mathcal{N}_{i\to j}^t$ indicates the total number of jumps $i\to j$ in a given time window of duration $t$. Denoting by
\begin{equation}
    \mathcal{P}[X] = \frac{\langle X^2 \rangle - \langle X \rangle^2}{\langle X \rangle^2}
\end{equation}
the squared coefficient of precision of a random variable $X$ with non-zero mean, the bound set by the TUR reads
\begin{equation}
    \mathcal{P}[\mathcal{N}_t] \geq \frac{2}{\sigma^{\rm ss} \, t} ,
    \label{eq:TUR}
\end{equation}
revealing that higher precision (i. e. narrower fluctuations) of $\mathcal{N}_t$ requires higher dissipation, quantified by $\sigma^{\rm ss}$, the stationary average rate of entropy production (here expressed in units of Boltzmann constant).

The TUR Eq.\,\eqref{eq:TUR} represents a principle of broad relevance for biochemical networks and molecular machines \cite{seifert_barato_TUR,liu2025dynamical,song2021thermodynamic,mallory2020we,gingrich2016dissipation,pietzonka2016universal,hwang2018energetic}. Extensions of this idea also regard bounds on multiple output currents \cite{dechant2018multidimensional} and on other performance metrics. For example, lower bounds have been derived for first-passage times \cite{falasco2020dissipation} and their fluctuations \cite{PhysRevResearch.3.L032034} in Markov processes, for the number of coherent oscillations (and timing precision) in biochemical clocks \cite{oberreiter2022universal,marsland2019thermodynamic}, and for information processing \cite{kamijima2025finite}. Similarly, the achievable signal-to-noise ratio of sensory readouts is fundamentally limited by dissipation \cite{liu2025dynamical}. In short, improving precision or speed in small systems inevitably incurs an irreducible thermodynamic cost.

Recent work has developed transition-based formulations of dissipation bounds when only partial information is available. In Ref.\,\cite{van2022thermodynamic} it is demonstrated that the log-ratio of waiting-time distributions for forward and backward transitions encodes the cycle affinity, yielding a lower-bound estimator for the total entropy production. Similarly, any partial estimate of the entropy production rate, derived from the statistics of visible transitions, provides a value lower than $\sigma^{\rm ss}$. These ideas have been extended into novel fluctuation relations. For instance, stopping an experiment after a fixed number of visible transitions recovers thermodynamic symmetry even in the presence of hidden dynamics \cite{singlecurrentFR}, and coarse-grained fluctuation theorems have been formulated for multiple observable currents \cite{Garilli_2024}. Universal trade-offs between entropy production and precision of counting observables or waiting times were also recently obtained \cite{PhysRevE.109.064128}.

In this context, we start from the known inequality \cite{PhysRevX.12.041026}
\begin{equation}
    \sigma^{\rm ss} \geq \sigma^\ell + \sigma^\tau \geq \sigma^\ell,
    \label{eq:known-eq}
\end{equation}
where $\sigma^\ell$ is a contribution that arises solely from the visible (observed) transitions on the monitored $\alpha\leftrightarrow\beta$, while $\sigma^\tau$ is a contribution that can be related to the hidden (not observed) part of the network. Then, $\sigma^\ell$ can be written as $\sigma^\ell = J \mathcal{A}$, where $J = \langle \mathcal{N}_t \rangle/t$ is the steady-state average probability current and $\mathcal{A}$ is an effective affinity. As shown in what follows, we provide an exact expression of $\sigma^\ell$ by further elaborating on a recently introduced explicit formula for the precision coefficient of an extensive current in a time window of duration $t$ \cite{cvexact}. By employing a decomposition of the moments of the recurrence times (i.e. the times between repeated occurrences of a given directed transition, possibly separated by an arbitrary number of opposite events) in terms of the moments of inter-transition times between visible transitions, we obtain an exact expression of $\lim_{t\to\infty} t\, \mathcal{P}[\mathcal{N}_t]$ in terms of $\mathcal{A}$ (and hence of $\sigma^\ell$) and of the precisions $\mathcal{P}[\tau_{\alpha\to\beta}]$, $\mathcal{P}[\tau_{\beta\to\alpha}]$ of the recurrence times of forward and backward transitions respectively. Exploiting this fact, and using the bound Eq.\,\eqref{eq:known-eq}, we provide a lower bound on the steady-state entropy production rate directly in terms of $\lim_{t\to\infty} t\, \mathcal{P}[\mathcal{N}_t]$, $\mathcal{P}[\tau_{\alpha\to\beta}]$ and $\mathcal{P}[\tau_{\beta\to\alpha}]$. This formulation, valid for any irreducible continuous-time Markov jump process, constrains the achievable trade-off between dissipation and experimentally accessible precisions, both of which represent key performance metrics of nanoscale systems, such as molecular motors or biochemical sensing networks \cite{liu2025dynamical}.
\section{Framework}

\def\side{1.8}
\def\spacing{0.2}

\begin{figure*}[ht!]
    \centering
    \includegraphics{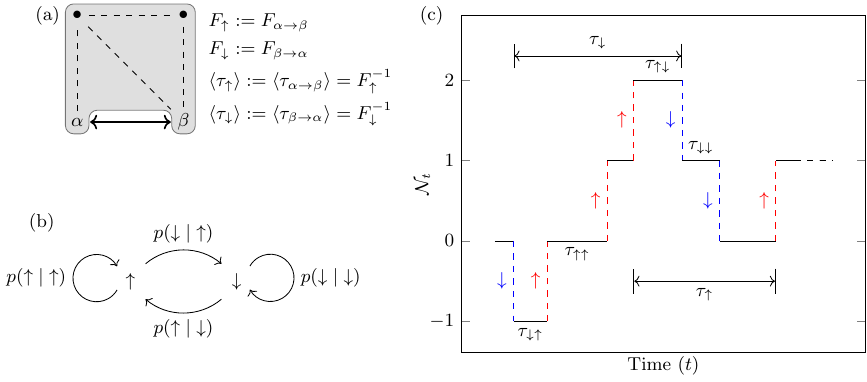}
    \caption{(a) An example of Markov jump process where only bidirectional transitions along a channel connecting states $\alpha$ and $\beta$ are observable. The shaded area represents the hidden part of the network. In this example, without loss of generality, only one channel $\alpha\leftrightarrow\beta$ is present; however, multiple channels connecting $\alpha$ and $\beta$ may exist, and they can be incorporated in the hidden part of the network, or can be lumped together in a single observable channel. (b) Network of the effective process in the space of transitions. The sequence of transitions $\ell \in \lbrace \up,\down\rbrace$, with $\up:\alpha \to \beta$ and $\down:\beta \to \alpha$, is generated from the discrete-time process with transition matrix $\boldsymbol{\rm P}$ [see Eq.\,\eqref{eq:trans-transition}]. (c) A typical realization of the process $\mathcal{N}_t$ where only $\up$ and $\down$ are observed. Specifically, a transition $\up$ increases the integrated current $\mathcal{N}_t$ by 1, whereas a transition $\down$ decreases the current by 1. Here, we also indicate the inter-transition times $\tau_{\ell'\ell}$ between consecutive occurrences of transitions $\ell'$ and $\ell$, and the recurrence times $\tau_{\ell}$ of transitions $\ell \in\lbrace \up,\down\rbrace$, which are blind to occurrences of the inverse transition $\bar{\ell}$.}
    \label{fig:setup}
\end{figure*}

We consider a Markov jump process on an irreducible network with a finite number of states. We assume, without loss of generality, that pairs of states are connected by a single bidirectional transition channel with a time-independent transition rate constant $k_{ij}$ from state $i$ to state $j$, and $k_{ji}$ from $j$ to $i$, ensuring that $\sigma^{\rm ss}$ is well defined at stationarity \cite{schnakenberg1976network}. The state space probability $\boldsymbol{\rm p}_t$ evolves according to the master equation $d\boldsymbol{\rm{p}}_t/dt = -\boldsymbol{\rm{R}} \boldsymbol{\rm{p}}_t $, with $[\boldsymbol{\rm{R}}]_{ij} = -k_{ji}(1-\delta_{ij}) + \delta_{ij}\sum_{n\neq i} k_{in}$ the rate matrix. For a pair of connected states $\alpha$ and $\beta$, we are interested in the extensive current
\begin{equation}
    \mathcal{N}_t = \mathcal{N}_{\alpha\to\beta}^t - \mathcal{N}_{\beta\to\alpha}^t,
\end{equation}
$\mathcal{N}_{i\to j}^t$ denoting the number of times the transition $i\to j$ occurs in a time window of duration $t$. Assuming that the system is initially found at stationarity, the integrated current $\mathcal{N}_t$ is on average $\langle \mathcal{N}_t \rangle = J t$ with $J$ the stationary probability current $J = {\rm p}
_\alpha^{\rm ss} k_{\alpha\beta} - {\rm p}_\beta^{\rm ss}k_{\beta\alpha} = F_{\alpha\beta} - F_{\beta\alpha}$, where $F_{ij} =  {\rm p}_i^{\rm ss} k_{ij}$ is the stationary flux from $i$ to $j$, and ${\rm p}_i^{\rm ss}$ is the stationary probability for state $i$.

An exact formula for the precision coefficient $\mathcal{P}[\mathcal{N}_t]$ of an extensive current $\mathcal{N}_t$ was recently presented in Ref.\,\cite{cvexact}. In particular, denoting with $\epsilon = J/\Omega$ the rectifying efficiency, with $\Omega = F_{\alpha\beta} + F_{\beta\alpha}$ the dynamical activity (or traffic) along the observed channel, and introducing the coefficient $c_0 = (1-\epsilon^2)/(2 \epsilon^2)$, for long observation times we have \cite{cvexact}
\begin{align}
    \mathcal{T}_\infty = \lim_{t\to\infty} t\;\mathcal{P}[\mathcal{N}_t] & = -\frac{1}{\epsilon J} + c_0 (\langle \tau_{\alpha\to\beta|\alpha} \rangle + \langle \tau_{\beta\to\alpha|\beta}\rangle)\nonumber \\
    &  \qquad + \frac{\mathcal{P}[\tau_{\alpha\to\beta}] - \mathcal{P}[\tau_{\beta\to\alpha}]}{J}.
    \label{eq:asymptotic-term}
\end{align}
where the coefficients
\begin{equation}
    \mathcal{P}[\tau_{i\to j}] = \frac{\langle \tau_{i\to j}^2\rangle -\langle \tau_{i\to j}\rangle^2}{\langle\tau_{i\to j}\rangle^2} ,
    \label{eq:precisions-time}
\end{equation}
quantify the precision of the recurrence times $\tau_{i\to j}$, whose mean values satisfy $\langle \tau_{i\to j}\rangle = F_{i j}^{-1}$. The quantities $\langle \tau_{i \to j | x_0}\rangle$ in Eq.\,\eqref{eq:asymptotic-term}, instead, represent the average occurrence times conditioned to the starting state $x_0$.

Equation\,\eqref{eq:asymptotic-term} represents the starting point to derive the main result [Eq.\,\eqref{eq:asymptotic-bound} later] through a recasting of $\langle \tau_{\alpha\to\beta|\alpha}\rangle + \langle \tau_{\beta\to\alpha|\beta}\rangle$ in terms of average recurrence times and ``trans-transition'' probabilities \cite{PhysRevX.12.041026,van2022thermodynamic,Garilli_2024,singlecurrentFR}. We briefly introduce the transition-based framework below.

\section{Visible transitions' dynamics}
To simplify the notation, let $\ell \in \lbrace \up,\down \rbrace$ denote the visible transition in its two possible directions, $\up: \alpha\to\beta$ and $\down: \beta \to \alpha$, and let $\mathtt{s}(\ell)$ and $\mathtt{t}(\ell)$ denote the source and target states of $\ell$ respectively. The inverse transition $\bar{\ell}$ is obtained by swapping the source and target states of $\ell$. Under the assumption that only the forward and backward jumps over the monitored channel are observable, the time series of the output consists of sequences of transitions $\ell \in \lbrace \up,\down \rbrace$ and the inter-transition times $\tau_{\ell'\ell}:=\tau_{\ell'\to\ell}$ [Fig.\,\ref{fig:setup}(c)].

It is a well established fact that sequences of single observable bidirectional transitions are generated by a discrete-time stochastic matrix $\boldsymbol{\rm P}$ whose elements are the trans-transition probabilities \cite{PhysRevX.12.041026,singlecurrentFR,Garilli_2024}
\begin{equation}
    [\boldsymbol{\rm P}]_{\ell\ell'} = p(\ell|\ell') = k_\ell [\boldsymbol{\rm S}^{-1}]_{\mathtt{s}(\ell)\mathtt{t}(\ell')} .
    \label{eq:trans-transition}
\end{equation}
The trans-transition probability $p(\ell|\ell')$ is the probability that the next observed transition is $\ell \in \lbrace \up,\down\rbrace$ given that the previous one was $\ell'\in \lbrace \up,\down\rbrace$, regardless of the time between the two occurrences, with $k_\ell \equiv k_{\mathtt{s}(\ell)\mathtt{t}(\ell)}$ and $p(\ell|\ell') + p(\bar{\ell}|\ell') = 1$ [see Fig.\,\ref{fig:setup}(b)]. The matrix $\boldsymbol{\rm S} := \boldsymbol{\rm R}^{\setminus \lbrace \up,\down \rbrace}$ \footnote{As explained in the main text, $\boldsymbol{\rm S}$ is obtained from the rate matrix $\boldsymbol{\rm R}$ by setting the elements $(\alpha,\beta)$ and $(\beta,\alpha)$ to zero, without modifying the diagonal entries. Therefore, throughout the paper, we indicate with the symbol $\setminus$ the removal of off-diagonal elements of a rate matrix associated to observable transitions. This should not be confused with stalling matrices \cite{polettini2017effective,polettini2019effective}, where the rates $k_{\alpha\beta}$ and $k_{\beta\alpha}$ are also removed from the diagonal of $\boldsymbol{\rm R}$}, also known as survival matrix, is obtained from the rate matrix $\boldsymbol{\rm R}$ as $[\boldsymbol{\rm S}]_{ij} = [\boldsymbol{\rm R}]_{ij}$ for all $i,j$ except for $[\boldsymbol{\rm S}]_{\alpha\beta} = [\boldsymbol{\rm S}]_{\beta\alpha} = 0$. Given the trans-transition probabilities Eq.\,\eqref{eq:trans-transition}, we can define the effective affinity, or force,
\begin{equation}
    \mathcal{A} = \ln \frac{p(\up|\up)}{p(\down|\down)} 
    \label{eq:effective-affinity}
\end{equation}
which drives the non-equilibrium current along the observed channel. The reduced effective entropy production rate mentioned in the Introduction,
\begin{equation}
    \sigma^\ell = J \mathcal{A},
    \label{eq:sigmaell}
\end{equation}
which is obtained solely from the observation of sequences of transitions regardless of the inter-transition times, lower bounds the stationary entropy production rate $\sigma^{\rm ss}$ of the full system, with equality holding for unicyclic networks \cite{PhysRevX.12.041026}. Additionally, the knowledge of the probability density of the inter-transition times allows the estimate of an extra entropic contribution, $\sigma^\tau$, related with transition-time asymmetries, providing the refined lower bound in Eq.\,\eqref{eq:known-eq}.

Interestingly, employing the transition-based formalism we can provide a simple relation between the average recurrence times and the average occurrence times conditioned to the source state, appearing in Eq.\,\eqref{eq:asymptotic-term}, via the trans-transition probabilities Eq.\,\eqref{eq:trans-transition}. In the following section, we use this fact to derive our results.

\section{Results}
The moments of the inter-transition times are obtained in terms of the survival matrix $\boldsymbol{\rm S}$ [see Appendix \ref{app:main-proof}]. The recurrence times $\tau_{\alpha\to \beta}$ and the conditional occurrence times $\tau_{\alpha \to \beta|x_0}$, on the other hand, are blind to the occurrences of the reverse transition $\beta\to \alpha$ between two occurrences of $\alpha\to \beta$ (the same holds for the inverse transition by swapping $\alpha$ and $\beta$). Therefore, they are expressed in terms of the survival matrix $\boldsymbol{\rm S}'  = \boldsymbol{\rm R}^{\setminus \lbrace \ell \rbrace}$ with $[\boldsymbol{\rm S}']_{ij} = [\boldsymbol{\rm R}]_{ij}$ and only $[\boldsymbol{\rm S}']_{\beta\alpha} = 0$.

Let $\tau_{\ell|x_0}$ denote the time at which $\ell \in\lbrace \up,\down\rbrace$ occurs conditioned to an initial state $x_0$. The recurrence time $\tau_\ell$ is obtained by setting $x_0 = \mathtt{t}(\ell)$. We prove in Appendix \ref{app:main-proof} that
\begin{equation}
    \langle \tau_{\ell|\mathtt{s}(\ell)}\rangle = \langle\tau_{\bar{\ell}\ell}\rangle + \frac{p(\bar{\ell}|\bar{\ell})}{p(\ell|\bar{\ell})} \langle\tau_{\bar{\ell}\bar{\ell}} \rangle
    \label{eq:occurrence-inter-transition}
\end{equation}
with $\tau_{\ell'\ell}$ the time between transition $\ell'$ (first) and transition $\ell$ (next). The average recurrence times thus satisfy
\begin{align}
    \langle \tau_\ell \rangle = p(\bar{\ell}|\ell) \left( \langle\tau_{\ell|\mathtt{s}(\ell)} \rangle + \langle\tau_{\bar{\ell}|\mathtt{s}(\bar{\ell})} \rangle \right),
    \label{eq:recurrence-occurrence}
\end{align}
which immediately provides
\begin{equation}
    \frac{\langle \tau_\ell \rangle}{\langle \tau_{\bar{\ell}} \rangle} = \frac{p(\bar{\ell}|\ell)}{p(\ell|\bar{\ell})} = \frac{F_{\bar{\ell}}}{F_{\ell}},
    \label{eq:local-fluxes}
\end{equation}
where we used $\langle\tau_\ell\rangle = F_\ell^{-1}$, being $F_{\ell}$ the stationary flux along the direction of $\ell$. Since the ratio between forward and backward fluxes is related with the local thermodynamic force $\mathcal{F}$, we straightforwardly obtain $\mathcal{F} = \ln\frac{p(\up|\down)}{p(\down|\up)}$, choosing $\ell = \up$ as the reference positive orientation of the transition. As a side outcome, we can show (see note Ref.\,\footnote{Let us introduce the ``flux matrix'' $\boldsymbol{\rm W} = \boldsymbol{\rm R} \boldsymbol{\rm \Pi}$, with $\boldsymbol{\rm R}$ the rate matrix and $\boldsymbol{\rm \Pi} = \text{diag}({\rm \boldsymbol{\rm p}^{\rm ss}})$. We have that $[\boldsymbol{\rm M}]_{ij} = - F_{ji}$ for $i \neq j$, and $[\boldsymbol{\rm W}]_{ii} = \sum_{n\neq i} F_{in}$. Thus, $\boldsymbol{\rm W}$ depends only on the full set of probability fluxes. Then, let $\boldsymbol{\tilde{\rm S}} = \boldsymbol{\rm S}\boldsymbol{\rm \Pi}$, from which we get $\boldsymbol{\rm S}^{-1} = \boldsymbol{\rm \Pi} \boldsymbol{\tilde{\rm S}}^{-1}$. The matrix $\boldsymbol{\tilde{\rm S}}$ is the analogue of a survival flux matrix built with $\boldsymbol{\rm W}$, hence $\boldsymbol{\tilde{\rm S}}^{-1}$ depends only on the full set of fluxes. By using Eq.\,\eqref{eq:trans-transition} we get $p(\ell|\ell') = k_\ell \sum_i [\boldsymbol{\rm W}]_{\mathtt{s}(\ell) i} [\boldsymbol{\tilde{\rm S}}^{-1}]_{i \mathtt{t}(\ell')} = k_\ell{\rm} {\rm p}_{\mathtt{s}(\ell)}^{\rm ss} [\boldsymbol{\tilde{\rm S}}^{-1}]_{\mathtt{s}(\ell) \mathtt{t}(\ell')} = F_\ell [\boldsymbol{\tilde{\rm S}}^{-1}]_{\mathtt{s}(\ell) \mathtt{t}(\ell')}$, which reveals that $p(\ell|\ell')$ depends only on the set of fluxes.}) that the trans-transition probabilities depend only on the set of probability fluxes. As a consequence, from Eq.\,\eqref{eq:recurrence-occurrence} we deduce that also the sum $\langle \tau_{\ell|\mathtt{s}(\ell)}\rangle + \langle\tau_{\bar{\ell}|\mathtt{s(\bar{\ell})}}\rangle$ is invariant for networks with the same set of fluxes, whereas the single addends are not.

To the best of our knowledge, Eqs.\,\eqref{eq:occurrence-inter-transition} to \eqref{eq:local-fluxes} are novel outcomes. In particular, Eq.\,\eqref{eq:recurrence-occurrence} represents a direct proportionality between average recurrence times and conditional occurrence times, the coefficient being expressed in terms of a trans-transition probability; Eq.\,\eqref{eq:local-fluxes} encodes the information of the local forces via the statistics of alternated series of visible transitions only.

Finally, for the pair of states $(\alpha,\beta)$, with $\up:=\alpha\to\beta$ and $\down:=\beta\to\alpha$ along the considered channel connecting $\alpha$ and $\beta$, let $J=F_\up - F_\down$ denote the stationary current. Using the expression Eq.\,\eqref{eq:recurrence-occurrence} we elaborate the term $\langle \tau_{\alpha\to\beta|\alpha}\rangle + \langle \tau_{\beta\to\alpha|\beta}\rangle$ in Eq.\,\eqref{eq:asymptotic-term}, obtaining, for $J \neq 0$,
\begin{align}
    \langle \tau_{\up|\alpha} \rangle+ \langle\tau_{\down|\beta}\rangle & = \frac{\langle \tau_\up\rangle - \langle\tau_\down\rangle}{p(\down|\up) - p(\up|\down)}\\
    & = \frac{\langle\tau_\up \rangle - \langle\tau_\down \rangle}{p(\down|\down) - p(\up|\up)},
    \label{eq:rewrite1}
\end{align}
where we used $p(\ell|\ell') = 1 - p(\bar{\ell}|\ell')$. As shown in Appendix\,\ref{app:eq16}, Eq.\,\eqref{eq:asymptotic-term} simplifies to
\begin{equation}
     \mathcal{T}_\infty = \frac{1}{J}\left(\frac{e^\mathcal{A} + 1}{e^\mathcal{A}-1} + \mathcal{P}[\tau_\up] - P[\tau_\down]\right),
     \label{eq:asymptnew}
\end{equation}
where we used the definition Eq.\,\eqref{eq:effective-affinity} for the effective affinity $\mathcal{A}$. Recalling the effective entropy production rate $\sigma^\ell = J \mathcal{A} \leq \sigma^{\rm ss}$ \cite{PhysRevX.12.041026}, we invert the expression above, obtaining our main result
\begin{equation}
    \sigma^{\rm ss} \geq \sigma^\ell = 2 J \coth^{-1} \left(J\mathcal{T}_\infty - \Delta \mathcal{P}^\tau\right),
    \label{eq:asymptotic-bound}
\end{equation}
with $\Delta \mathcal{P}^\tau = \mathcal{P}[\tau_\up] - \mathcal{P}[\tau_\down]$. The quantity $J\mathcal{T}_\infty$ is also known as Fano factor \cite{schnitzer1995statistical,moffitt2014extracting,moffitt2010methods}. For comparison, the asymptotic TUR from Eq.\,\eqref{eq:TUR} reads
\begin{equation}
    \sigma^{\rm ss} \geq \sigma^{\rm TUR} = \frac{2}{\mathcal{T}_\infty}.
    \label{eq:asymptotic-TUR}
\end{equation}

The bound in Eq.\,\eqref{eq:asymptotic-bound} is meaningful whenever $J$ is nonzero, while for vanishing current it reduces to the trivial inequality $\sigma^{\rm ss} \ge 0$. Remarkably, the inequality \eqref{eq:asymptotic-bound} surely saturates for unicyclic networks because $\sigma^\ell = \sigma^{\rm ss}$ in such a special case \cite{PhysRevX.12.041026,van2022thermodynamic}. Note that, when $J=0$, it can be shown that [see Appendix\,\eqref{app:localequilibrium}]
\begin{equation}
    \Delta \mathcal{P}^\tau_{J=0} = 2 F \frac{1-p}{p}\left(\langle \tau_{\down\down}\rangle - \langle \tau_{\up\up}\rangle\right) ,
    \label{eq:localequilibrium}
\end{equation}
where $F = F_\up = F_\down$ and $p = p(\up|\down)_{J=0} = p(\down|\up)_{J=0}$, which means that $\Delta \mathcal{P}^\tau_{J=0}$ can assume finite values on local equilibrium conditions, highlighting the presence of hidden dissipation. Only in the unicyclic case, $J=0$ implies $\Delta \mathcal{P}^\tau_{J=0} = 0$, since $\langle \tau_{\up\up}\rangle = \langle \tau_{\down\down}\rangle$ \cite{PhysRevX.12.041026}. Conversely, $J$ can be non-null even if $\Delta \mathcal{P}^\tau = 0$. Additionally, Eq.\,\eqref{eq:asymptotic-bound} yields the bound $\vert J\mathcal{T}_\infty - \Delta \mathcal{P}^\tau \vert > 1$ because $\coth^{-1}(x)$ is only defined for $\vert x \vert > 1$.

The expression Eq.\,\eqref{eq:asymptotic-bound} reveals that a lower bound on the stationary entropy production rate $\sigma^{\rm ss}$ of the full system is obtained after evaluating the factor $\mathcal{T}_\infty = \lim_{t\to\infty} t\;\mathcal{P}[\mathcal{N}_t]$ at long times, and collecting the statistics of the recurrence times $\tau_{\alpha\to\beta}$ and $\tau_{\beta\to\alpha}$. It is worth noting that, at least in the unicyclic case \cite{wierenga2018quantifying}, the estimate of $\mathcal{T}_\infty$ does not necessarily require sampling infinitely long trajectories. In such case, in fact, $\mathcal{T}_\infty$ can be inferred from finite-time measurements through the statistics of the times between strictly positive increments of the observed current \cite{wierenga2018quantifying}.

\begin{figure}
    \centering
    \includegraphics[width=.90\linewidth]{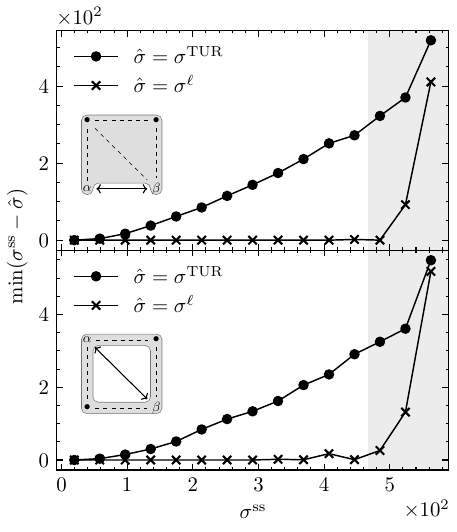}
    \caption{Comparison between lower bounds for the entropy production rate $\sigma^{\rm ss}$ obtained from the long-time precision coefficient of a single integrated current: asymptotic TUR Eq. \eqref{eq:asymptotic-TUR} (circles) vs. asymptotic bound Eq. \eqref{eq:asymptotic-bound} (crosses). The points are obtained from randomly generated instances of the 4-state network in Fig.\,\ref{fig:setup}(a), with rate constants drawn as $k = 10^X$, where $X$ is uniformly distributed in the interval $[-1,2)$. The values of $\sigma^{\rm ss}$ are divided into equally sized intervals. Then, for each interval, we pick the minimum value of the difference $\sigma^{\rm ss} - \hat{\sigma}$, for $\hat{\sigma} = \sigma^{\rm TUR},\sigma^\ell$. For the same network, each panel represents a different choice of observable edge, highlighted by the bidirectional arrow in the insets. The irregularities at very high $\sigma^{\rm ss}$ are due to the limited amount of data falling in that region (less than 10 networks per interval, shaded region).}
    \label{fig:comparison-asymptotic}
\end{figure}

As a numerical check [see Fig.\,\ref{fig:comparison-asymptotic}], we performed extensive tests on randomly generated networks to confirm that $\sigma^\ell$ computed with Eq.\,\eqref{eq:asymptotic-bound} never falls above the true entropy production rate $\sigma^{\rm ss}$. In addition, we found that this bound can provide almost saturating estimates of $\sigma^{\rm ss}$ under less stringent requirements than the TUR. For randomly generated rate constants, in fact, Eq.\,\eqref{eq:asymptotic-bound} can still approach saturation in the far-from-equilibrium regime, even for networks that are not (quasi)-unicyclic; for an example, see the note Ref.\,\footnote{From the randomly-generated networks used for Fig.\,\ref{fig:comparison-asymptotic}, it is possible to find some examples where the bound Eq.\,\eqref{eq:asymptotic-bound} is close to saturation despite not being unicyclic. For instance, let us enumerate the sites clockwise starting from the right-bottom one, and assign rates $k_{12} = 2,\, k_{21} = 78,\,k_{13} =  40,\, k_{31} = 0.25,\, k_{14} = 4,\, k_{41} = 0.4,\, k_{23} = 0.25,\, k_{32} = 47,\, k_{34} = 2$ and $k_{43} = 40$. This network is not unicyclic and is found to be far from equilibrium ($\sigma^{\rm ss} \sim 230$). When observed along the channel $\alpha\leftrightarrow \beta$, with $\alpha = 2$ and $\beta = 1$, it provides $(\sigma^{\rm ss} - \sigma^\ell)/\sigma^{\rm ss} \sim 10^{-4}$, so that it can be considered very close to saturation. For a comparison, the same set of rate constant provides the estimate $(\sigma^{\rm ss} - \sigma^{\rm TUR})/\sigma^{\rm ss} \sim 0.62$ when using the asymptotic TUR.}.

\section{Discussion and final remarks}

The bound $\sigma^\ell \leq \sigma^{\rm ss}$ was originally formulated in terms of stalling probabilities \cite{polettini2017effective,polettini2019effective,raghu2025effective} or from the statistics of sequences of visible transitions \cite{PhysRevX.12.041026,singlecurrentFR}, and it only accounts for asymmetries of extensive current-like quantities. Here we have rephrased this inequality in terms of directly measurable fluctuation statistics. Specifically, our main result, Eq.\,\eqref{eq:asymptotic-bound}, involves the long-time precisions on a chosen integrated steady current $\mathcal{N}_t$ and on the forward and backward recurrence times associated with the corresponding transitions. This reformulation makes the connection between entropy production rate and measurable precision parameters explicit, thereby providing a TUR-like inequality where, besides $\mathcal{T}_\infty$, also $\Delta \mathcal{P}^\tau$ does enter. It shows that one can infer a minimal steady-state entropy production rate from the asymptotic variability of the integrated current $\mathcal{N}_t$ and the asymmetry in the recurrence-times fluctuations. 

Interestingly, our inequality can approach saturation even for randomly generated networks far from equilibrium, where the TUR typically becomes loose. In particular, equality in Eq.\,\eqref{eq:asymptotic-bound} holds for any unicyclic network with arbitrary rate constants, but there are non-unicyclic networks for which the bound is practically saturated \footnotemark[\value{footnote}]. The TUR, instead, achieves saturation only under special conditions, such as in the linear response regime or for unicyclic networks with homogeneous rates \cite{seifert_barato_TUR}.

Remarkably, the lower bound $\sigma^\ell$ considered in this work is not directly comparable to the TUR, and an universal order relation does not exist between the two formulations, except when $\Delta \mathcal{P}^\tau = 0$ \,\footnote{For the very special case where $\Delta \mathcal{P}^\tau = 0$ and $\vert J \mathcal{T}_\infty \vert > 1$, the bound Eq.\,\eqref{eq:asymptotic-bound} is always tighter than the TUR, since $2 J \coth^{-1}(J\mathcal{T}_\infty) > 2/\mathcal{T}_\infty$.}.

The structure of inequality \eqref{eq:asymptotic-bound} also opens a route for optimization. Since it couples precision on the integrated current $\mathcal{N}_t$, recurrence-time precision, speed (i.e. the average current $J$), and dissipation rate, it provides a natural framework to study trade-offs among these quantities. For example, one may tune the rate constants to enhance the output's precision at the expense of higher energetic cost, or analyze the efficiency of molecular clocks and timers in terms of the balance between accuracy and dissipation. Such applications highlight how fluctuation-based bounds can go beyond inference and might serve as design principles for nonequilibrium processes.

Because the bound Eq.\,\eqref{eq:asymptotic-bound} is expressed in terms of observable fluctuations, it provides a tangible route to quantify dissipation in complex systems where affinities or full state information remain hidden. The practical implications are broad. In biophysics, for instance, steady currents and recurrence-time distributions are often experimentally accessible, thanks to single-molecule tracking techniques, especially for rotary molecular motors (such as the F$_1$-ATPase motor \cite{yasuda1998f1,watanabe2008effect,PhysRevLett.104.218103,mishima2025efficiently}) and transporters \cite{svoboda1993direct,coppin1997load,nishiyama2002chemomechanical,carter2005mechanics,pietzonka2016universal,wirth2023minflux,deguchi2023direct}. Furthermore, the randomness of catalytic sites in single-molecule enzymology is central in the field of statistical kinetics \cite{schnitzer1995statistical,moffitt2014extracting,moffitt2010methods}. Systems of this type constitute potential test benches of our results, provided that  an $\alpha\leftrightarrow\beta$ channel be identifiable and observable (e.g. associated with stepping events in processive motors, or with reactive events of product formation in enzymatic catalysis), a condition that is explicitly invoked in some theoretical models \cite{liepelt2007kinesin,kolomeisky2007molecular}.

Looking forward, a particularly promising direction is to explore more deeply the information contained in forward and backward recurrence-time statistics. Asymmetries between these distributions encode irreversibility through their Kullback–Leibler divergence, potentially providing a direct microscopic link to entropy production. Here, we have shown that this information is already captured by the difference $\Delta \mathcal{P}^\tau$ between precision coefficients, suggesting a systematic way to detect dissipation even when average currents vanish. A detailed analysis of these temporal asymmetries beyond the second moments of the distributions might therefore sharpen the present bound and clarify the mechanisms by which hidden dynamics contribute to entropy production.

Overall, our contribution lies in rephrasing a known transition-based entropy production bound [Eq.\,\eqref{eq:known-eq}] in terms of measurable precision coefficients, thereby clarifying the role of both currents and recurrence-time fluctuations. This perspective not only strengthens thermodynamic inference but also highlights new opportunities for optimization and design, underscoring the relevance of precision-based bounds for the study of nonequilibrium processes in biological and chemical systems.

\section{Acknowledgements}
The authors acknowledge the financial contribution from “Fondazione Cassa di Risparmio di Padova e Rovigo” (CARIPARO) within the framework of the project ``NoneQ'', ID 68058.

\bibliography{biblio}

\appendix

\onecolumngrid

\section{Proof of Eqs.\,\eqref{eq:occurrence-inter-transition} and \eqref{eq:recurrence-occurrence} }
\label{app:main-proof}

Let $\up: \alpha\to \beta$ be the transition from $\alpha$ to $\beta$ through the (chosen) channel connecting $\alpha$ and $\beta$, and let $\down:\beta \to\alpha$ be its reversed counterpart. Also let $\boldsymbol{\rm S}':=\boldsymbol{\rm R}^{\setminus \lbrace \ell \rbrace},\,\ell \in \lbrace \up,\down \rbrace$, indicate the survival matrix derived for the first-exit-time problem where the given transition $\ell$ is absorbant, and let $\boldsymbol{\rm S} := \boldsymbol{\rm R}^{\setminus \lbrace \up,\down \rbrace}$ indicate the survival matrix where both $\up$ and $\down$ are absorbant.

According to our sign convention on rate matrices, the average conditional occurrence time for transition $\ell$ is \cite{frezzato2020stationary,cvexact}
\begin{equation}
    \langle \tau_{\ell|x_0}\rangle = k_\ell\left[\left(\boldsymbol{\rm S}'\right)^{-2}\right]_{\mathtt{s}(\ell) x_0} = \sum_n \left[\left(\boldsymbol{\rm S}'\right)^{-1}\right]_{n x_0}.
\end{equation}
Notice that the reverse transition $\bar{\ell}$ can occur arbitrarily many times before $\ell$ occurs. 

Here we show that the average occurrence times can be expressed as a linear combination of the inter-transition times which are obtained from a first-exit problem where both $\up$ and $\down$ are absorbant, and therefore depend on the survival matrix $\boldsymbol{\rm S}$. The four average inter-transition times $\langle \tau_{\ell'\ell} \rangle$ between consecutive transitions $\ell'$ and $\ell$ are \cite{PhysRevX.12.041026}
\begin{equation}
    \langle \tau_{\ell'\ell} \rangle = \frac{[\boldsymbol{\rm S}^{-2}]_{\mathtt{s}(\ell)\mathtt{t}(\ell')}}{[\boldsymbol{\rm S}^{-1}]_{\mathtt{s}(\ell)\mathtt{t}(\ell')}}.
    \label{eq:inter-transition}
\end{equation}

Given an initial state $x_0$ at the initial time $t_0 = 0$, we are interested in the probability density $\rho(\tau,\ell|x_0)$ associated with the occurrence of $\ell$ in the time interval $[\tau,\tau+d\tau)$. Letting $\mathcal{N}(\bar{\ell})$ indicate the number of times $\bar{\ell}$ occurs before the occurrence of $\ell$ given $x_0$, the density $\rho(\tau,\ell|x_0)$ can be decomposed as
\begin{equation}
    \rho(\tau,\ell |x_0) = \sum_{n=0}^\infty \rho(\tau,\ell;\lbrace \mathcal{N}(\bar{\ell}) = n  \rbrace | x_0)
    \label{eq:recurrence-density}
\end{equation}
where
\begin{equation}
    \rho(\tau,\ell;\lbrace \mathcal{N}(\bar{\ell}) = n \rbrace | x_0) = \int_{0}^\tau dt_n \int_0^{t_n} dt_{n-1}\int_0^{t_{n-1}} dt_{n-2} \dots \int_{0}^{t_2} dt_1\; \rho(\tau,\ell; t_n, \bar{\ell} ; t_{n-1}, \bar{\ell};\cdots ; t_1, \bar{\ell} | x_0)
    \label{eq:convolution}
\end{equation}
is interpreted as a convolution of probability densities of the kind $[e^{-(t_i - t_{i-1}) \boldsymbol{\rm S} }]_{xy_0} k_{xy}$, for suitable choices of the states $x$, $y$ and $y_0$ [see below]. In what follows we treat separately the cases where $x_0 = \mathtt{s}(\ell)$ and $x_0 = \mathtt{t}(\ell)$, and we solve the problem by means of the Laplace transform $\mathcal{L}\lbrace \rho \rbrace$.

\subsection{Case 1: $x_0 = \mathtt{s}(\ell)$}
Let us consider the case $x_0 = \mathtt{s}(\ell)$. The integrand in Eq.\,\eqref{eq:convolution} becomes, for each $n$,
\begin{equation}
    \rho(\tau,\ell; t_n, \bar{\ell} ; t_{n-1}, \bar{\ell};\cdots ; t_1, \bar{\ell} | \mathtt{s}(\ell)) = {k_\ell}\left[e^{-(\tau-t_n)\boldsymbol{\rm S}}\right]_{\mathtt{s}(\ell)\mathtt{s}(\ell)} \left(k_{\bar{\ell}}\right)^n \left(\prod_{i = 1}^n \left[e^{-(t_i-t_{i-1})\boldsymbol{\rm S}}\right]_{\mathtt{s}(\bar{\ell})\mathtt{s}(\ell)}\right),
    \label{eq:integrand}
\end{equation}
with $t_0 = 0$. The Laplace transform of $e^{-t \boldsymbol{\rm S}}$ reads
\begin{equation}
    \int_0^\infty dt\, e^{-t u} e^{-t \boldsymbol{\rm S}} = \left(u\boldsymbol{\rm I} + \boldsymbol{\rm S}\right)^{-1} = \boldsymbol{\rm S}(u)^{-1},
    \label{eq:laplace-single}
\end{equation}
with $\boldsymbol{\rm I}$ the identity matrix, and where we defined $\boldsymbol{\rm S}(u) = u\boldsymbol{\rm I} + \boldsymbol{\rm S}$, which satisfies $\boldsymbol{\rm S}(0) = \boldsymbol{\rm S}$ and $\frac{d}{du}\boldsymbol{\rm S}(-u)^{-1} = \boldsymbol{\rm S}(-u)^{-2}$. The Laplace transform of Eq.\,\eqref{eq:recurrence-density}, by considering Eq.\,\eqref{eq:integrand} along with Eq.\,\eqref{eq:laplace-single}, is then
\begin{equation}
    \mathcal{L}\lbrace \rho(\tau,\ell|\mathtt{s}(\ell) )\rbrace (u) = {k_\ell} [\boldsymbol{\rm S}(u)^{-1}]_{\mathtt{s}(\ell)\mathtt{s(\ell)}} \sum_{n=0}^\infty   \left( k_{\bar{\ell}} [\boldsymbol{\rm S}(u)^{-1}]_{\mathtt{s}(\bar{\ell})\mathtt{s(\ell)}}\right)^n
    \label{eq:laplace-occurrence}
\end{equation}
By taking the derivative $\left(d \mathcal{L}(-u)/du\right)_{u = 0}$ we obtain the average occurrence times Eq.\,\eqref{eq:occurrence-inter-transition} conditioned to the source state of $\ell$:
\begin{align}
    \langle \tau_{\ell|\mathtt{s}(\ell)}\rangle &= \left(\frac{k_\ell [\boldsymbol{\rm S}(-u)^{-2}]_{\mathtt{s}(\ell)\mathtt{s}(\ell)}}{1-{k_{\bar{\ell}}}[\boldsymbol{\rm S}(-u)^{-1}]_{\mathtt{s}(\bar{\ell})\mathtt{s}(\ell)}}\right)_{u=0} + \lim_{M\to\infty}\left(k_\ell [\boldsymbol{\rm S}(-u)^{-1}]_{\mathtt{s}(\ell)\mathtt{s}(\ell)}  \frac{d}{du} \frac{1-\left[{k_{\bar{\ell}}}[\boldsymbol{\rm S}(-u)^{-1}]_{\mathtt{s}(\bar{\ell})\mathtt{s}(\ell)}\right]^{M+1}}{1-{k_{\bar{\ell}}}[\boldsymbol{\rm S}(-u)^{-1}]_{\mathtt{s}(\bar{\ell})\mathtt{s}(\ell)}}\right)_{u=0}\\
    & = {\frac{{k_\ell} [\boldsymbol{\rm S}^{-2}]_{\mathtt{s}(\ell)\mathtt{s}(\ell)}}{1-{k_{\bar{\ell}}}[\boldsymbol{\rm S}^{-1}]_{\mathtt{s}(\bar{\ell})\mathtt{s}(\ell)}}} + k_\ell [\boldsymbol{\rm S}^{-1}]_{\mathtt{s}(\ell)\mathtt{s}(\ell)}\left(\frac{{k_{\bar{\ell}}}[\boldsymbol{\rm S}^{-2}]_{\mathtt{s}(\bar{\ell})\mathtt{s}(\ell)}}{\left(1-{k_{\bar{\ell}}}[\boldsymbol{\rm S}^{-1}]_{\mathtt{s}(\bar{\ell})\mathtt{s}(\ell)}\right)^2}\right),
\end{align}
where we used the partial sum for the geometric series and then summed after taking $u = 0$. Finally, by considering that $\mathtt{s}(\ell) = \mathtt{t}(\bar{\ell})$ and by recognizing the inter-transition times Eq.\,\eqref{eq:inter-transition} and the trans-transition probabilities Eq.\,\eqref{eq:trans-transition}, we obtain Eq.\,\eqref{eq:occurrence-inter-transition}:
\begin{equation}
    \langle \tau_{\ell|\mathtt{s}(\ell)}\rangle = \langle \tau_{\bar{\ell}\ell}\rangle + \frac{p(\bar{\ell}|\bar{\ell})}{p(\ell|\bar{\ell})}\langle\tau_{\bar{\ell}\bar{\ell}}\rangle .
   \label{eq:app-occurrence}
\end{equation}

\subsection{Case 2: $x_0 = \mathtt{t}(\ell)$}
Similarly to the previous case, the Laplace transform of Eq.\,\ref{eq:recurrence-density} for $x_0 = \mathtt{t}(\ell) = \mathtt{s}(\bar{\ell})$ reads
\begin{equation}
    \mathcal{L}\lbrace \rho(t,\ell|\mathtt{t}(\ell)) \rbrace(u) = {k_\ell}[\boldsymbol{\rm S}(u)^{-1}]_{\mathtt{s}(\ell)\mathtt{s}(\bar{\ell})} + {k_\ell} {k_{\bar{\ell}}}[\boldsymbol{\rm S}(u)^{-1}]_{\mathtt{s}(\ell)\mathtt{s}(\ell)} [\boldsymbol{\rm S}(u)^{-1}]_{\mathtt{s}(\bar{\ell})\mathtt{s}(\bar{\ell})} \sum_{n = 0}^\infty \left( {k_{\bar{\ell}}} [\boldsymbol{\rm S}(u)^{-1}]_{\mathtt{s}(\bar{\ell})\mathtt{s}(\ell)}\right)^n 
    \label{eq:laplace-recurrence}
\end{equation}
Taking $\left(d \mathcal{L}(-u)/du\right)_{u = 0}$ and using similar arguments to those used to get Eq.\,\eqref{eq:app-occurrence}, we obtain the average occurrence times conditioned to the target state of $\ell$, which is exactly the average recurrence time $\langle \tau_{\ell} \rangle$ of transition $\ell$. This corresponds to Eq.\,\eqref{eq:recurrence-occurrence}:
\begin{equation}
    \langle \tau_{\ell}\rangle = p(\bar{\ell}|\ell) \left(\langle \tau_{\ell\bar{\ell}}\rangle + \langle \tau_{\bar{\ell}\ell}\rangle + \frac{p(\ell|\ell)}{p(\bar{\ell}|\ell)} \langle \tau_{\ell\ell}\rangle + \frac{p(\bar{\ell}|\bar{\ell})}{p(\ell|\bar{\ell})} \langle \tau_{\bar{\ell}\bar{\ell}}\rangle\right)  = p(\bar{\ell}|\ell) \left( \langle \tau_{\ell|\mathtt{s}(\ell)}\rangle + \langle \tau_{\bar{\ell}|\mathtt{s}(\bar{\ell})}\rangle \right).
    \label{eq:app-recurrence}
\end{equation}

\section{Proof of Eq.\,\eqref{eq:asymptnew}}
\label{app:eq16}

Given the stochastic matrix Eq.\,\eqref{eq:trans-transition}, the steady state vector of the transition-space process, $\boldsymbol{\rm q}^{\rm ss} = ({\rm q}_\up^{\rm ss},{\rm q}_\down^{\rm ss})^T$, is such that $\boldsymbol{\rm q}^{\rm ss} = \boldsymbol{\rm P}\boldsymbol{\rm q}^{\rm ss}$, with $\boldsymbol{\rm P}$ defined in Eq.\,\eqref{eq:trans-transition}. It can be easily verified that ${\rm q}_\up^{\rm ss} = p(\up|\down)/(p(\up|\down) + p(\down|\up))$ and ${\rm q}_\down^{\rm ss} = p(\down|\up)/(p(\up|\down) + p(\down|\up))$, from which we obtain the direct relation between the rectifying efficiency $\epsilon$ and $\boldsymbol{\rm q}^{\rm ss}$:
\begin{equation}
    \epsilon = {\rm q}^{\rm ss}(\up) - {\rm q}^{\rm ss}(\down) = \frac{p(\up|\down) - p(\down|\up)}{p(\up|\down) + p(\down|\up)}.
\end{equation}
Since the factor $c_0$ entering Eq.\,\eqref{eq:asymptotic-term} can be expressed as
\begin{equation}
    c_0 = \frac{2}{J}\frac{1}{\langle \tau_\down \rangle - \langle \tau_\up \rangle},
\end{equation}
by using Eq.\,\eqref{eq:rewrite1} we obtain that 
\begin{equation}
    -\frac{1}{\epsilon J} + c_0 (\langle \tau_{\up|\alpha}\rangle + \langle \tau_{\down|\beta}\rangle) = \frac{1}{J} \left( \frac{p(\up|\up) + p(\down|\down)}{p(\up|\up) - p(\down|\down)} \right) = \frac{1}{J}\left(\frac{e^{\mathcal{A}}+1}{e^{\mathcal{A}}-1}\right),
\end{equation}
where we also used $p(\ell|\ell') = 1- p(\bar{\ell}|\ell')$ and Eq.\,\eqref{eq:effective-affinity}. Plugging the expression above into Eq.\,\eqref{eq:asymptotic-term}, we finally obtain Eq.\,\eqref{eq:asymptnew}.

\section{$\Delta \mathcal{P}^\tau$ at local equilibrium}
\label{app:localequilibrium}

Here we derive the exact expression Eq.\,\eqref{eq:localequilibrium}, valid in local equilibrium conditions. Very generally, $\Delta\mathcal{P}^\tau$ can be written as
\begin{equation}
    \Delta \mathcal{P}^\tau = \frac{\langle \tau_{\up}^2 \rangle - \langle \tau_{\up}\rangle^2}{\langle \tau_\up\rangle^2} - \frac{\langle \tau_{\down}^2 \rangle - \langle \tau_{\down}\rangle^2}{\langle \tau_\down\rangle^2} = F_{\up}^2 \langle \tau_{\up}^2\rangle - F_{\down}^2 \langle \tau_{\down}^2\rangle,
    \label{eq:rewritingDeltaP}
\end{equation}
where we used that $\langle \tau_\ell\rangle = F_\ell^{-1}$. We explicitly evaluate the second moments of the recurrence times $\tau_\ell$ by employing the same techniques used for the first moments in Appendix\,\ref{app:main-proof}. Taking the second derivative $\left(d^2 \mathcal{L}(-u)/du^2\right)_{u = 0}$ of Eq.\,\eqref{eq:laplace-recurrence}, we obtain the general expression of $\langle \tau_\ell^2 \rangle$
\begin{align}
    \langle\tau_\ell^2 \rangle = & p(\bar{\ell}|\ell) \left( y + z(\ell)\right) ,
    \label{eq:secondmoments-recurrence}
\end{align}
in terms of the moments of the inter-transition times $\tau_{\ell'\ell}$, with $\ell,\ell' \in \lbrace \up,\down \rbrace$, where
\begin{equation}
    y = \langle\tau_{\down\up}^2\rangle + \langle\tau_{\up\down}^2\rangle + 2\langle\tau_{\up\down}\rangle \langle\tau_{\down\up}\rangle + \frac{p(\up|\up)}{p(\down|\up)}\langle\tau_{\up\up}^2\rangle  + \frac{p(\down|\down)}{p(\up|\down)}\langle\tau_{\down\down}^2\rangle,
\end{equation}
is independent of $\ell$, and 
\begin{equation}
    z(\ell) = 2\frac{p(\bar{\ell}|\bar{\ell})}{p(\ell|\bar{\ell})}\langle \tau_{\bar{\ell}\bar{\ell}}\rangle(\langle\tau_{\up\down}\rangle + \langle\tau_{\down\up}\rangle) + 2\left(\frac{p(\bar{\ell}|\bar{\ell})}{p(\ell|\bar{\ell})}\right)^2 \langle\tau_{\bar{\ell}\bar{\ell}}\rangle^2.
    \label{eq:zeta}
\end{equation}
In the above expressions, $\langle \tau_{\ell'\ell}^2 \rangle = 2[\boldsymbol{\rm S}^{-3}]_{\mathtt{s}(\ell)\mathtt{t}(\ell')}/[\boldsymbol{\rm S}^{-1}]_{\mathtt{s}(\ell)\mathtt{t}(\ell')}$. Before proceeding, we need to prove the identity
\begin{equation}
    \langle \tau_{\up\down} \rangle + \langle \tau_{\down\up}\rangle = \frac{p(\up|\down) + p(\down|\up)}{p(\up|\down)p(\down|\up)\,\Omega} - \left(\frac{p(\up|\up)}{p(\down|\up)}\right)\langle \tau_{\up\up}\rangle - \left(\frac{p(\down|\down)}{p(\up|\down)}\right) \langle \tau_{\down\down}\rangle,
    \label{eq:auxiliary-relation}
\end{equation}
with $\Omega = F_\up + F_\down$ the local dynamical activity along the observed channel. Let us consider a time window of length $t$, where the sequence of $n$ bidirectional visible transitions $\lbrace \ell_1,\dots,\ell_n \rbrace$ is observed, with $\ell_i$ either $\up$ or $\down$. The time $t$ can be decomposed as
\begin{equation}
    t = t(\ell_1) + (t(\ell_2)-t(\ell_1)) + \cdots + (t-t(\ell_n)),
\end{equation}
with $t(\ell_i)$ the time at which transition $\ell_i$ occurs. Since $t(\ell_i) - t(\ell_{i-1})$ is the inter-transition time $\tau_{\ell_{i-1}\ell_i}$, we write
\begin{equation}
    t = t(\ell_1) + (t-t(\ell_n)) + \sum_{\ell,\ell' \in \lbrace \up,\down \rbrace}\sum_{i = 1}^{n_{\ell'\ell}} \tau_{\ell'\ell}^{(i)},
\end{equation}
where $\tau_{\ell'\ell}^{(i)}$ denotes the inter-transition time of the $i$-th occurrence of $\ell' \to \ell$ and $n_{\ell'\ell}$ the total number of $\ell'\to\ell$ transitions in a trajectory of duration $t$. Dividing the expression above by $t$ and taking the limit $t\to\infty$, the boundary terms $t(\ell_1)/t$ and $(t-t(\ell_n))/t$ become negligible, and therefore
\begin{equation}
    \lim_{t\to\infty} \frac{1}{t} \sum_{\ell,\ell' \in \lbrace \up,\down \rbrace}\sum_{i = 1}^{n_{\ell'\ell}} \tau_{\ell'\ell}^{(i)} = \lim_{t\to\infty}\frac{1}{t} \sum_{\ell,\ell' \in \lbrace \up,\down \rbrace} n_{\ell'\ell} \langle \tau_{\ell'\ell} \rangle = \Omega \sum_{\ell,\ell' \in \lbrace \up,\down \rbrace} {\rm q}^{\rm ss} (\ell') p(\ell|\ell') \langle \tau_{\ell'\ell} \rangle = 1,
\end{equation}
where we identified $\langle \tau_{\ell'\ell}\rangle$ with the empirical average $\hat{\tau}_{\ell'\ell} = \sum_{i = 1}^{n_{\ell'\ell}} \tau_{\ell'\ell}^{(i)}/n_{\ell'\ell}$, and used $\lim_{t\to\infty} n_{\ell'\ell}/t = \Omega \, {\rm q}^{\rm ss}(\ell')p(\ell|\ell')$. Using ${\rm q}_\up^{\rm ss} = p(\up|\down)/(p(\up|\down) + p(\down|\up))$ and ${\rm q}_\down^{\rm ss} = p(\down|\up)/(p(\up|\down) + p(\down|\up))$, we obtain
\begin{equation}
    \frac{\Omega}{p(\up|\down)+p(\down|\up)} \left[p(\up|\down)p(\down|\up) \left( \langle\tau_{\up\down} \rangle + \langle \tau_{\down\up} \rangle \right) + p(\up|\down)p(\up|\up)\langle \tau_{\up\up}\rangle + p(\down|\up) p(\down|\down) \langle \tau_{\down\down}\rangle \right] = 1,
\end{equation}
which, by inversion, provides Eq.\,\eqref{eq:auxiliary-relation}. 

When $J=0$, then $F_\up = F_\down = F$ and also $\Omega = 2F$. Since $J = \Omega \epsilon$ vanishes if and only if $\epsilon = 0$, then the steady-state probabilities for the discrete-time process in the space of transitions (see Appendix\,\ref{app:eq16}) satisfy ${\rm q^{\rm ss}} (\up) = {\rm q^{\rm ss}} (\down) = 1/2$, implying that $p(\up|\down) = p(\down|\up) = p$ and therefore $p(\up|\up) = p(\down|\down) = 1-p$. By plugging Eq.\,\eqref{eq:auxiliary-relation} into Eq.\,\eqref{eq:zeta} and evaluating at $J=0$, we obtain
\begin{equation}
    z(\ell)_{J=0} = 2\frac{1-p}{p}\left(\frac{\langle\tau_{\bar{\ell}\bar{\ell}}\rangle}{p F} - \frac{1-p}{p} \langle \tau_{\up\up}\rangle \langle \tau_{\down\down}\rangle\right).
    \label{eq:zetaell}
\end{equation}
For $J=0$, Eq.\,\eqref{eq:rewritingDeltaP} reduces to
\begin{equation}
    \Delta \mathcal{P}^\tau_{J=0} = pF^2\left[ z(\up)_{J=0} -  z(\down)_{J=0} \right],
    \label{eq:deltaP-local-eq}
\end{equation}
which, using Eq.\,\eqref{eq:zetaell}, readily provides Eq.\,\eqref{eq:localequilibrium} in the main text.

\end{document}